\documentclass{article}
\usepackage{sao1}
\usepackage{graphicx}
\usepackage{amssymb}

\DeclareTextSymbol{\degre}{T1}{6}
\DeclareTextSymbol{\degre}{OT1}{23}

\begin{document}

\title{The MiMeS Project: Overview and current status}

\author{G.A. Wade and the MiMeS Collaboration\thanks{www.physics.queensu.ca/$\sim$wade/mimes}}

\institute{Department of Physics, Royal Military College of Canada, PO Box 17000 Station Forces, Kingston, Ontario, Canada, K7K 7B4}    

\maketitle

\begin{abstract}
The Magnetism in Massive Stars (MiMeS) Project is a consensus collaboration among many of the foremost international researchers
of the physics of hot, massive stars, with the basic aim of understanding the origin, evolution and impact of 
magnetic fields in these objects. At the time of writing, MiMeS Large Programs have acquired over 1250 high-resolution polarised spectra of about 150 individual stars with spectral types from B5-O4,
discovering new magnetic fields in over a dozen hot, massive stars. Notable results include the detection of magnetic fields in the two most rapidly-rotating known magnetic stars, and in the most massive known magnetic star. In this paper we review the structure of the MiMeS observing programs and report the status of observations, data modeling and development of related theory, and review important results obtained so far.
\end{abstract}

\section{Introduction}

Massive stars are those stars with initial masses above about 8 times that of the sun, eventually ending their lives in core-collapse (types Ib, Ic and II) supernovae. These represent the most massive and luminous stellar component of the Universe, and are the crucibles in which the lion's share of the chemical elements are forged. These rapidly-evolving stars drive the chemistry, structure and evolution of galaxies, dominating the ecology of the Universe - not only as supernovae, but also during their entire lifetimes - with far-reaching consequences. 

The Magnetism in Massive Stars (MiMeS) Project represents a comprehensive, multidisciplinary strategy to address the Òbig questionsÓ related to the complex and puzzling magnetism of massive stars. MiMeS has been awarded "Large Program" status by the Canada-France-Hawaii Telescope (CFHT) and the T\'elescope Bernard Lyot (TBL), resulting in a total of 1230 hours of time allocated to the Project with the high-resolution spectropolarimeters ESPaDOnS and Narval from late 2008 through 2012. This commitment of the observatories, their staff, their resources and expertise is being used to acquire an immense database of sensitive measurements of the optical spectra and magnetic fields of massive stars, which will be combined with a wealth of new and archival complementary data (e.g. optical photometry, UV and X-ray spectroscopy), and applied to address the 4 main scientific objectives of the MiMeS Project: 

\begin{enumerate}
\item The origin and evolution physics of magnetic fields in massive stars: The weight of opinion holds that the magnetic fields of massive stars are fossil fields - the slowly-decaying remnants of magnetic field accumulated or generated during the complex process of star formation. Although the fossil paradigm provides a powerful framework for interpreting the magnetic characteristics of massive stars, its physical details are only just beginning to be elaborated.  Significant work remains in order to fully explore and test the fossil field hypothesis, and to understand the detailed physics leading from the microgauss fields in the interstellar medium to the kilogauss fields observed in main sequence stars. 

\item The physics of atmospheres, winds, envelopes and magnetospheres of hot stars: The supersonic, radiatively driven winds of massive stars couple with their rapid rotation and strong magnetic fields, generating complex and dynamic magnetospheric structures whose observational signatures span the entire electromagnetic spectrum. MiMeS collaborators have developed sophisticated multi-D numerical models that can furnish detailed, quantitative predictions of the structure and dynamics of massive-star magnetospheres, and as the theoretical models become progressively refined, there are growing demands for high-quality observational data that the MiMeS project furnishes -- optical, ultraviolet and X-ray spectroscopy, precision multi-colour photometry, radio measurements, and detailed magnetic field topologies and abundance maps --  for the models' validation, calibration and exploration. 

\item The rotation and rotational evolution of massive stars: The interaction of the stellar winds of hot, massive stars with magnetic fields modifies mass loss, and may enhance the shedding of rotational angular momentum via magnetic braking. Among the intermediate-mass magnetic stars, the influence of magnetic braking (likely occurring during pre-main sequence evolution) is obvious - their rotational velocities are on average 3 times slower than non-magnetic single stars of similar mass. Whereas magnetic massive stars are likely subjected to weaker pre-main sequence braking and possibly even spin-up, the opportunity exists for significant braking to occur on the main sequence via their strong winds. Consequently, in massive stars the impact of magnetic fields on rotation is much less clear. 

\item The evolution of massive stars and origin of the magnetic fields of neutron stars and magnetars: The evolutionary life cycles of hot, massive stars may be strongly modified by the presence of a surface magnetic field. In addition to the important impact of surface magnetic fields on rotation and mass loss, magnetic fields inside the star influence internal differential rotation and circulation currents. The relationship of these internal magnetic fields to the strong ($10^{11}-10^{15}$~G) fields of neutron stars and magnetars, which represent the ultimate endpoint of the evolution of many massive stars, is currently a topic of vigorous debate. Are neutron star magnetic fields fossils, or are they generated during the violent processes accompanying their formation in supernovae? And how are they related to ÒhypernovaeÓ and gamma-ray bursts? Whatever the origin of these fields, the magnetic and physical characteristics of their main sequence and post-main sequence progenitors (e.g. massive supergiants and Wolf-Rayet stars) that are being observed by MiMeS represent fundamental constraints and input for supernova simulations, neutron star field generation theories and stellar evolution models.  
\end{enumerate}

\section{Steering committee}

The MiMeS Collaboration has grown to over 50 collaborators from nearly 30 institutions in Canada, France and around the world. The project is overseen and directed by a steering committee comprised of French, Canadian and international experts:
\\

\begin{tabular}{ll}
Overall coordination:       &   	G.A. Wade (RMC, Canada) [Project and CFHT LP PI]\\
Be stars:                               &   	C. Neiner (LESIA, France) [TBL LP PI]\\
Field O stars:                      &   	J.-C. Bouret (LAM, France)\\
Field B stars:                     &    	H. Henrichs (Amsterdam, Netherlands)\\
PMS HBe stars:                &  	E. Alecian (LAOG/LESIA, France)\\
Cluster OB stars:              &  	V. Petit (West Chester, USA)\\
WR Stars:                          &    	N. St. Louis (MontrŽal, Canada)\\
Data Archiving:                 &  	D. Bohlender (NRC Victoria, Canada)\\
Modeling:                           &   	R. Townsend (Wisconsin), O. Kochukhov (Uppsala, Sweden)\\
Theory:                                &  	 S. Mathis (Saclay, France)\\
Phase 2 coordination:       & 	 J. Grunhut (RMC, Canada)\\
\end{tabular}

\section{Infrastructure and support}

MiMeS operates a public www site ({\tt www.physics.queensu.ca/$\sim$wade/mimes}) that describes the project and its scientific objectives, and reports the most recent scientific results. In addition, the Canadian Astronomy Data Centre maintains the MiMeS Data Archive (www.cadc-ccda.hia-iha.nrc-cnrc.gc.ca/MiMeS), which provides secured access to MiMeS CFHT data in raw and reduced form. A MiMeS Wiki site provides a private forum for collaboration discussion, sharing of data, results and tools, etc. Finally, a dedicated MiMeS server (located in Paris, and managed by C. Neiner) is employed by the collaboration for data processing and analysis, and as an repository and backup of all MiMeS data.
Project activities are supported financially through operating grants of various collaborators in North America, and French PNPS funding to support travel of French MiMeS members, and to support meetings and visits to France. 

\section{Periodic workshops and related meetings}

MiMeS workshops have been held at least annually since the beginning of the project in 2008:
\\

\begin{tabular}{l}
 1st MiMeS Workshop: Kingston, Canada, July 2008\\
 2nd MiMeS Workshop: Paris, France, May 2009\\
 3rd MiMeS Workshop: Waimea, USA, November 2009\\
 4th MiMeS Workshop: Armagh, Northern Ireland, July 2010\\
 \end{tabular}
 \\
 
 The next MiMeS workshop is planned for June 2011 in Madison, USA. Visit the MiMeS www site\footnote{www.physics.queensu.ca/wade/$\sim$mimes} for details.
 
 In addition, MiMeS collaborators have been responsible for the organisation of major international conferences aligned with MiMeS objectives. These include the Core Collapse to YSOs Workshop in London, Canada in May 2010, and IAUS 272: Active OB Stars in Paris, France in July 2010.

\section{Structure of the Large Programs}

\begin{figure}
\center
 \includegraphics[width=12.5cm]{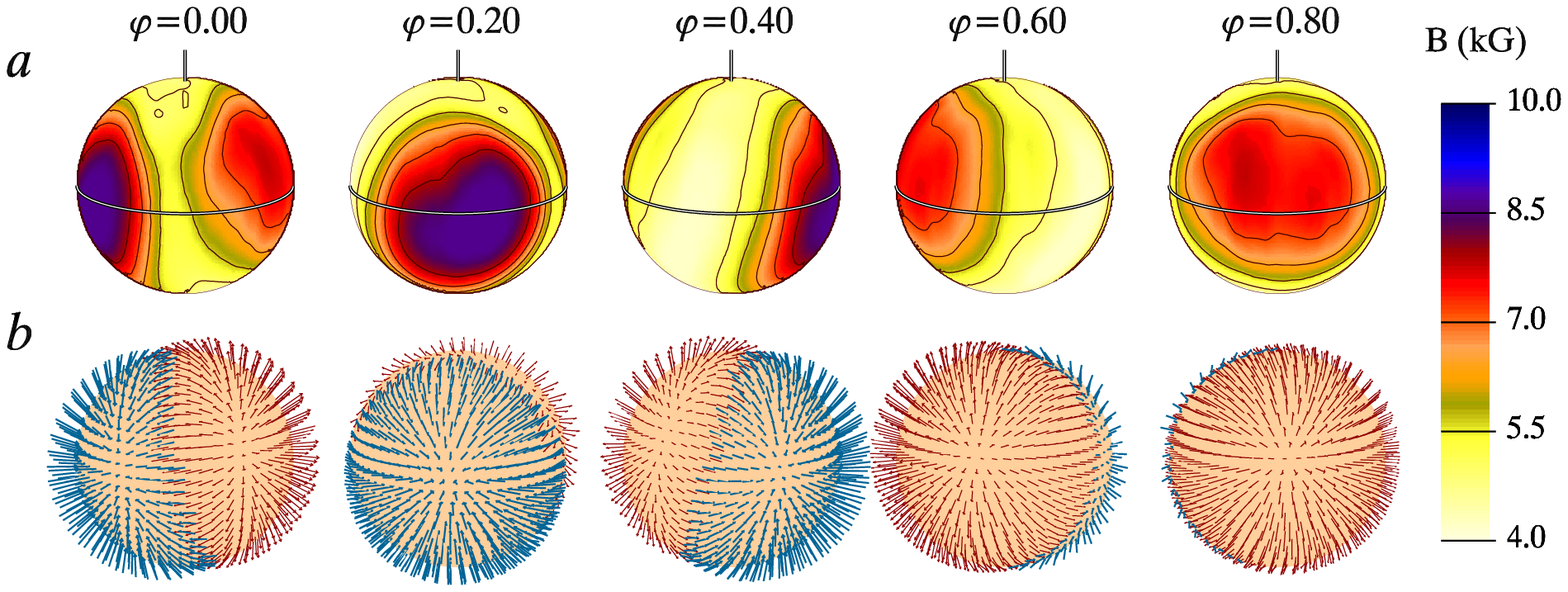}
 \includegraphics[width=12.5cm]{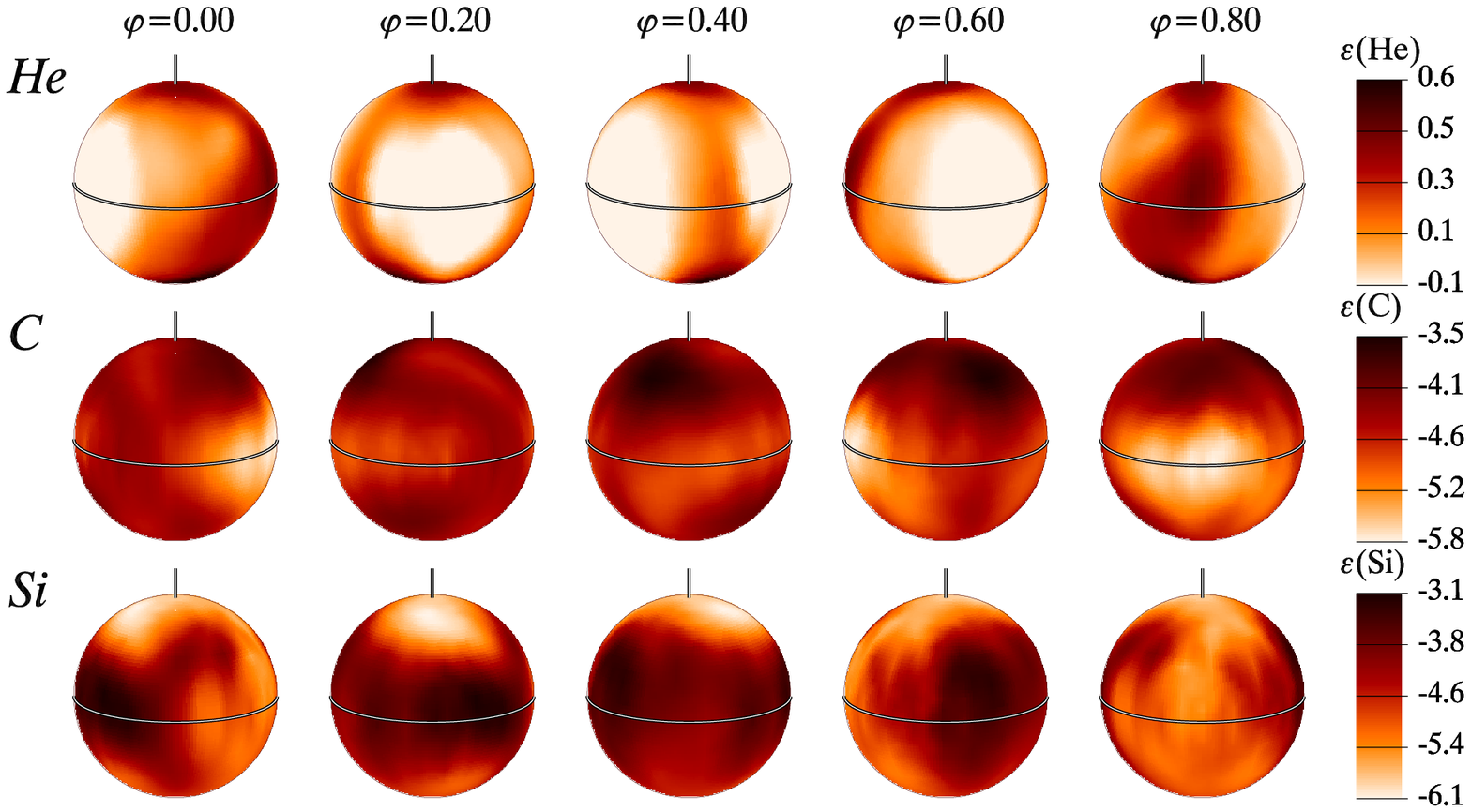}
  \caption{Magnetic Doppler Imaging (MDI) of the B2Vp star $\sigma$~Ori E (HD 37479; Oksala et al., in prep), illustrating the reconstructed magnetic field intensity (upper row) and orientation (lower row) and the abundance distributions of He, C and Si of this star at 5 equidistant rotation phases. The maps were obtained from a time-series of 18 Stokes $IV$ spectral sequences. Maps similar to these will be constructed for other stars in the MiMeS Targeted Component.}\label{fig:wave}
\end{figure}

To address the general problems expressed by the MiMeS science drivers, we have devised a two-component observing program that 
will allow us to obtain basic statistical information about the magnetic properties of the overall 
population of hot, massive stars (the Survey Component), while simultaneously providing detailed 
information about the magnetic fields and related physics of individual objects (the Targeted Component).

\subsection{Targeted Component} 

The MiMeS Targeted Component (TC) will provide data to map the magnetic fields and 
investigate the physical characteristics of a sample of known magnetic stars of great interest, 
at the highest level of sophistication possible. The roughly 25 TC targets have been selected to allow us to investigate a variety of 
physical phenomena, and to allow us to directly and quantitatively confront the predictions of stellar evolution theory, 
MHD magnetised wind simulations, magnetic braking models, etc. 

Each TC target is to be observed many times with ESPaDOnS and Narval, in order to obtain a high-precision and
high-resolution sampling of the rotationally-modulated circular (and sometimes linear) polarisation line profiles. Using state-of-the-art
tomographic reconstruction techniques such as Magnetic Doppler Imaging (Piskunov \& Kochukhov 2002), detailed maps
of the vector magnetic field on and above the surface of the star will be constructed (e.g. see Fig. 1). In combination with new and archival
complementary data, detailed analyses will be undertaken to model their evolutionary states, rotational evolution and wind structure and dynamics.

\begin{figure}
\center
 \includegraphics[width=5.3cm]{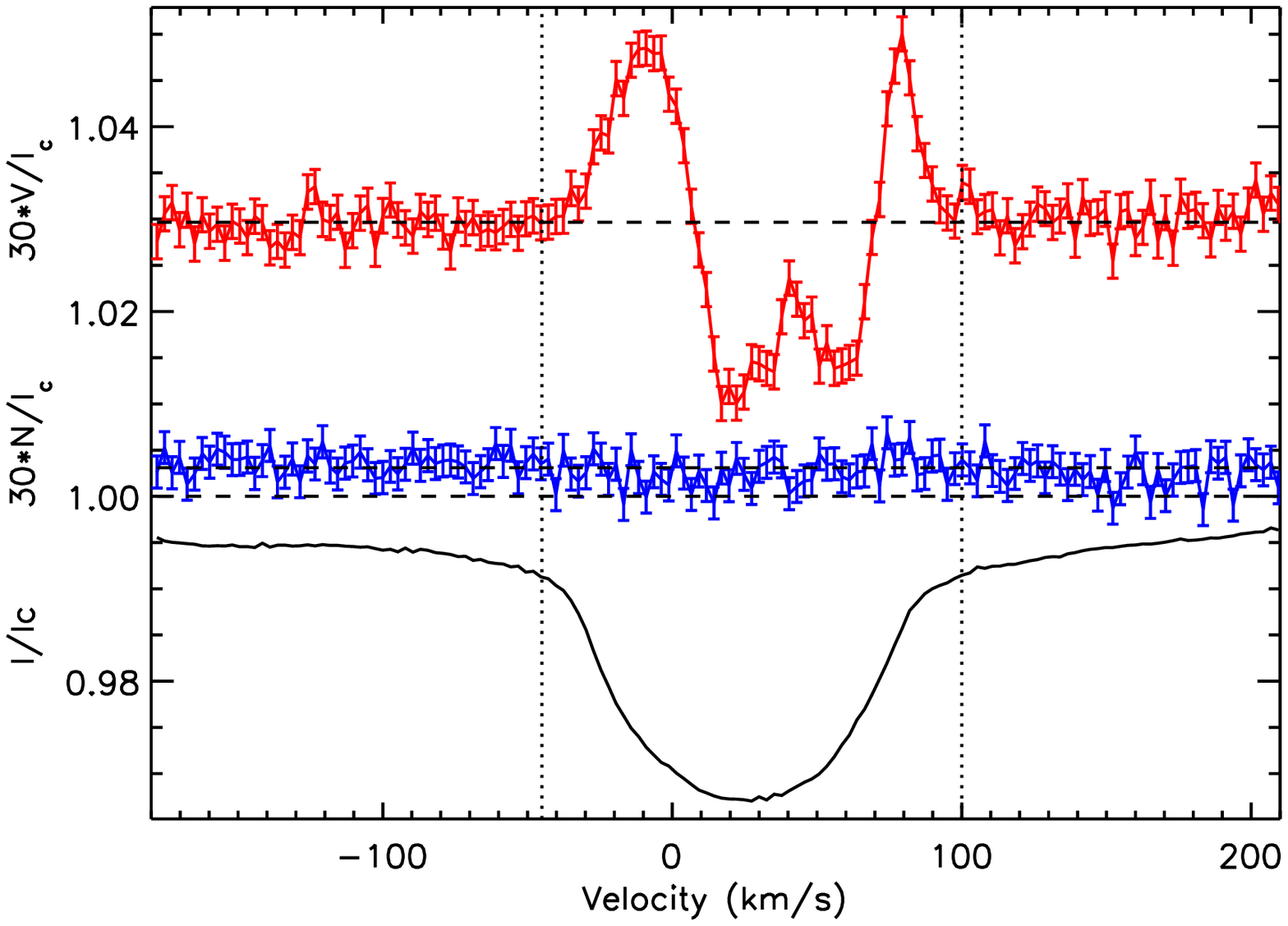}  \includegraphics[width=5.3cm]{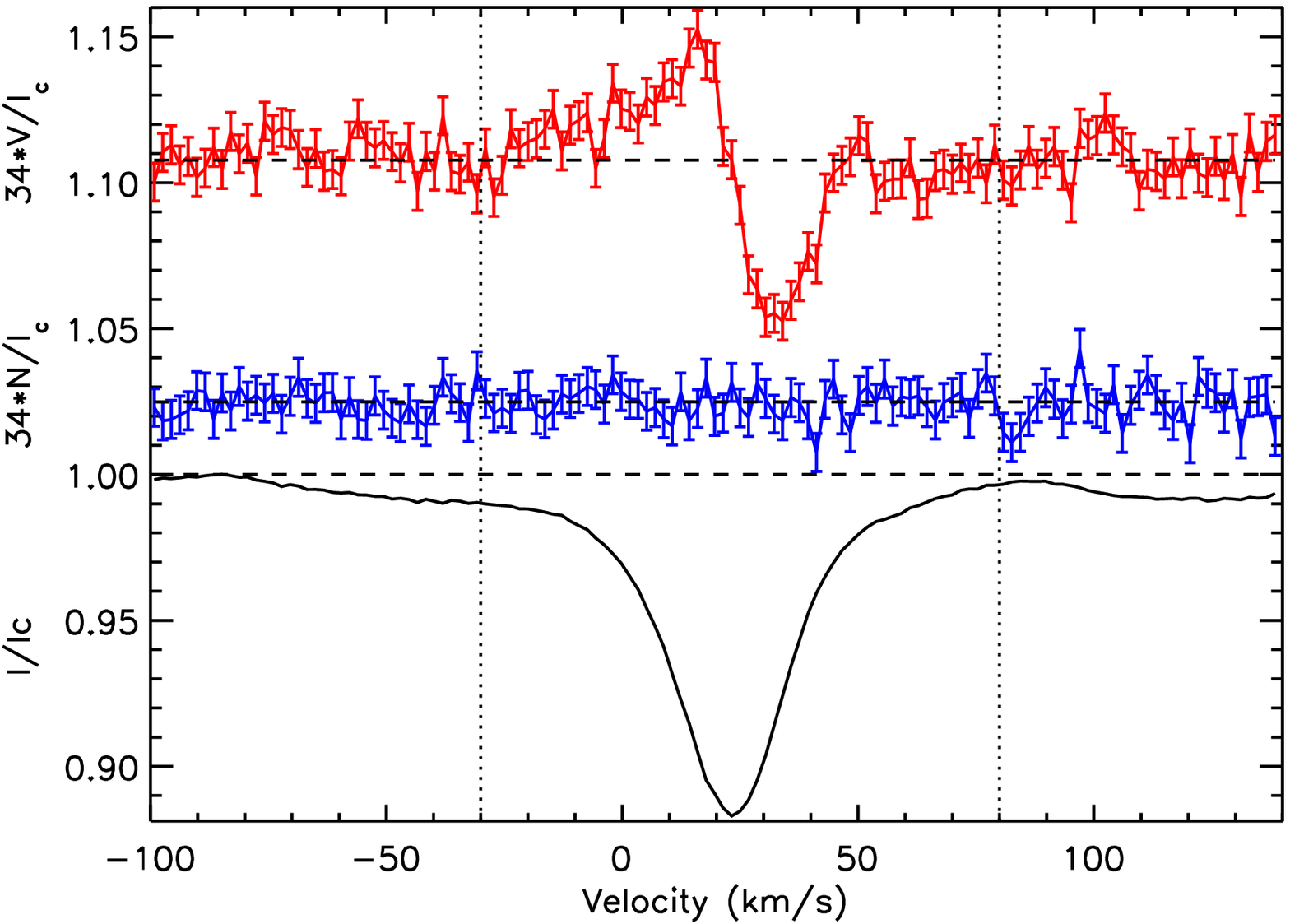} \includegraphics[width=5.3cm]{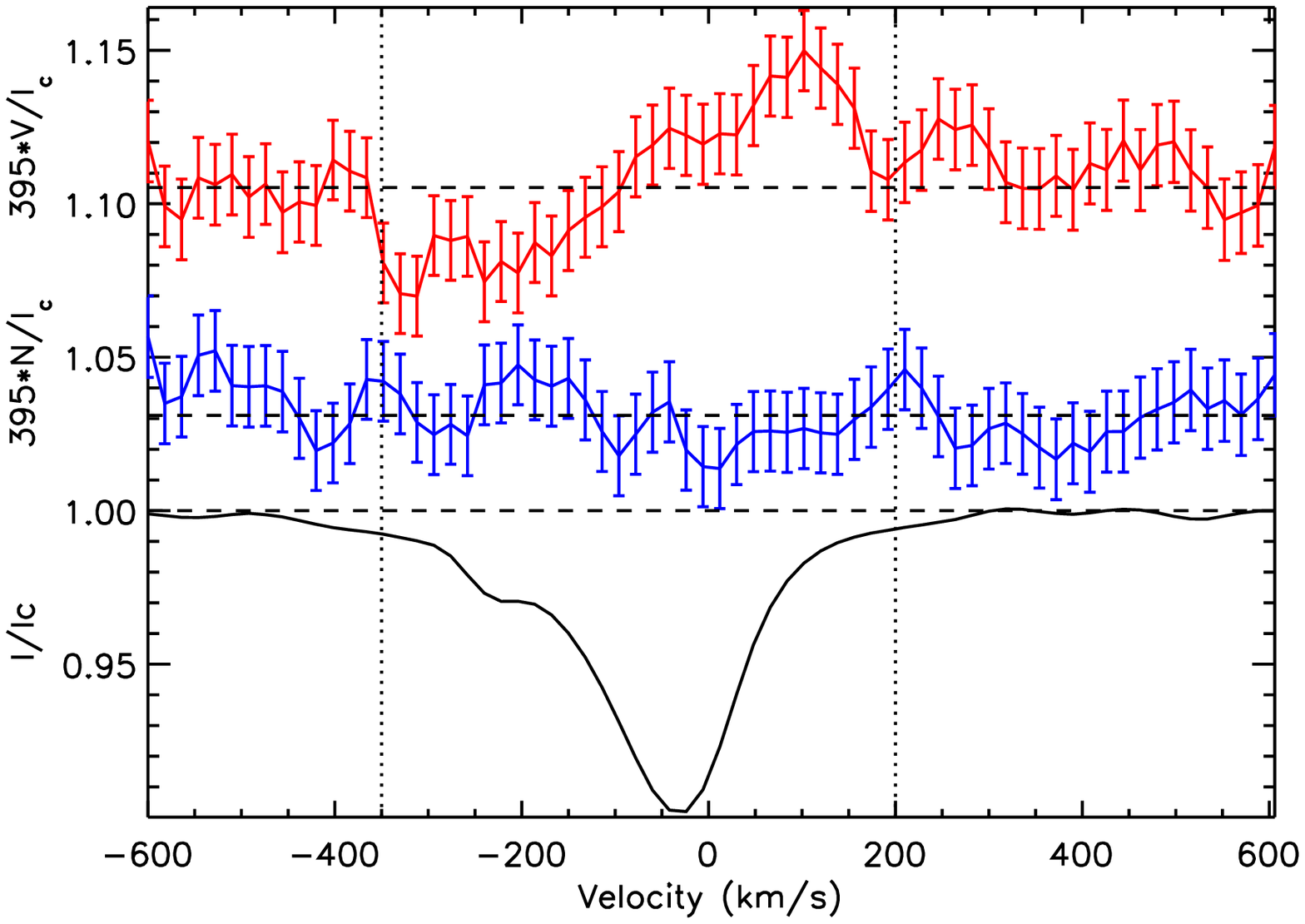}
  \caption{Least-Squares Deconvolved profiles of 3 hot stars in which magnetic fields have been discovered by the MiMeS Project: HD~61556 (B5V, left, Rivinius et al. 2010), HD 57682 (O9IV, middle, Grunhut et al. 2009 and these proceedings) and HD 148937 (Of?p, right, Wade et al., in prep). The curves show the mean Stokes $I$ profiles (bottom curve), the mean Stokes $V$ profiles (top curve) and the $N$ diagnostic null profiles (middle curve). Each star exhibits a clear magnetic signature in Stokes $V$. To date, a dozen new magnetic stars have been discovered through the MiMeS Survey Component. }\label{fig:wave}
\end{figure}

\subsection{Survey Component} 

The MiMeS Survey Component (SC) provides critical missing information about field 
incidence and statistical field properties for a much larger sample of massive stars. It will also serve to 
provide a broader physical context for interpretation of the results of the Targeted Component.  
From an extensive list of potential OB stars compiled from published catalogues, we have 
generated an SC target sample of about 200 stars, covering the full range of spectral types from B4-O4, which 
are selected to sample broadly the parameter space of interest, while being well-suited to field detection. Our target list includes pre-main sequence Herbig Be stars,
field and cluster OB stars, Be stars, and Wolf-Rayet stars. 

Each SC target has been, or will be, observed once or twice during the Project, at very high precision in circular polarisation (e.g. see Figs. 2 \& 3). From the SC data we will
measure the bulk incidence of magnetic massive stars, estimate the variation of incidence versus 
mass, derive the statistical properties (intensity and geometry) of the magnetic fields of massive stars, 
estimate the dependence of incidence on age and environment, and derive the general statistical 
relationships between magnetic field characteristics and X-ray emission, wind properties, rotation, 
variability, binarity and surface chemistry diagnostics.  

Of the 1230 hours allocated to the MiMeS LPs, about one-half is assigned to the TC and one-half to the SC. 

\section{Precision magnetometry of massive stars}

\begin{figure}
\center
 \includegraphics[width=15cm]{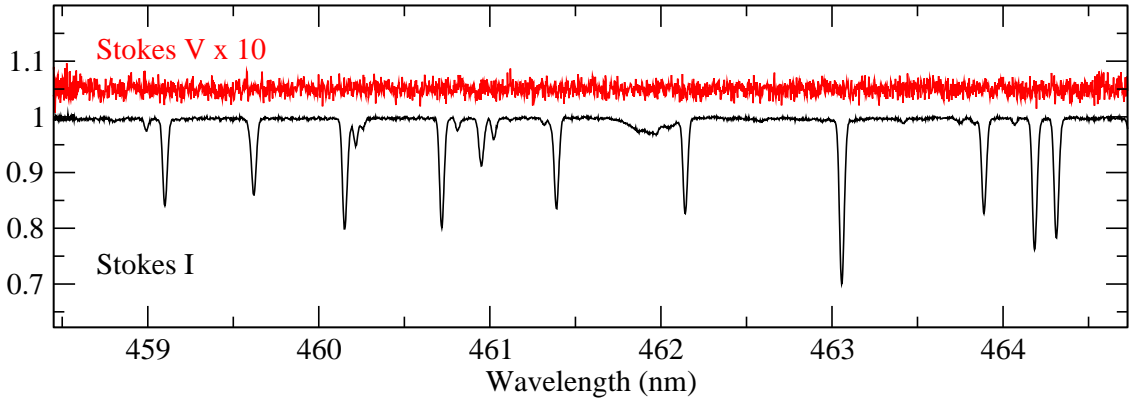}
  \caption{Detail of the MiMeS SC spectrum of the sharp-lined $\beta$~Cep star $\delta$~Ceti (= HD 16582). The peak S/N per 1.8 km/s pixel is 1000 (for an exposure time of 280 s), typical for an SC observation. LSD analysis yields no evidence of a magnetic field, with a 1$\sigma$ longitudinal field error bar of just 10~G.}\label{fig:wave}
\end{figure}

For all targets we exploit the longitudinal Zeeman effect in metal and helium lines to detect and measure magnetic 
fields in the line-forming region. Splitting of a spectral line due to a longitudinal magnetic field into oppositely polarised $\sigma$ components produces a variation of circular polarisation across the line (commonly referred to as a Ò(Stokes $V$) Zeeman signatureÓ or Òmagnetic signatureÓ; see Fig. 2.). The amplitude and morphology of the Zeeman signature encode information about the strength and structure of the global magnetic field. 
For some TC targets, we will also exploit the transverse Zeeman effect to constrain the detailed local structure of the field. Splitting of a spectral line by a transverse magnetic field into 
oppositely polarised $\pi$ and $\sigma$ components produces a variation of linear polarisation (characterized by 
the Stokes $Q$ and $U$ parameters) across the line (e.g. Kochukhov et al. 2004, Petit 2010).

\subsection{Survey Component}

For the SC targets, the detection of magnetic field is diagnosed using the Stokes $V$ detection criterion described by Donati et al. (1997), and the probability densities and associated "odds ratio" computed using the powerful Bayesian 
estimation technique of Petit et al. (2008). After reduction of the polarised spectra using the Libre-Esprit optimal extraction code (see Fig. 3), we employ the Least-Squares Deconvolution (LSD; Donati et al. 1997) multi-line 
analysis procedure to combine the Stokes $V$ Zeeman signatures from many spectral lines into a single high-S/N mean profile (see Fig. 2), enhancing our ability to detect subtle magnetic signatures. Least-Squares 
Deconvolution of a spectrum requires a Òline maskÓ to describe the positions, relative strengths and magnetic sensitivities of the lines predicted to occur in the stellar spectrum.  In our analysis we employ custom line masks that we tailor interactively to best reproduce the observed stellar spectrum, in order to maximise our sensitivity to weak magnetic fields.

The exposure duration required to detect a Zeeman signature of a given strength 
varies as a function of stellar apparent magnitude, spectral type and projected rotational velocity. This 
results in a large range of detection sensitivities for our targets. The SC exposure times are based on 
an empirical exposure time relation derived from real ESPaDOnS observations of OB stars, and takes into account detection sensitivity gains resulting from LSD and velocity 
binning, and sensitivity losses from line broadening due to rapid rotation.  
Exposure times for our SC targets correspond to the time required to 
definitely detect (with a false alarm probability below $10^{-5}$) the Stokes $V$ Zeeman signature produced 
by a surface dipole magnetic field with a specified polar intensity from 0.1-1 kG. Although our calculated exposure times correspond to definite detections of a dipole magnetic field, 
our observations are also sensitive to substantially more complex field topologies.

Upon detection of a magnetic field in an SC target, the star is transferred to the TC and becomes the subject of intensive monitoring to confirm the field, to detect variability of the spectrum and polarisation profiles, and ultimately to determine the stellar rotational period, magnetic field characteristics and associated stellar properties. Examples of such discoveries are HD 57682 (Grunhut et al. 2009 and these proceedings) and HR 7355 (Oksala et al. 2010, Sect. 8.2, and Wade et al. in these proceedings).

A few examples of SC targets in which magnetic fields have been discovered by MiMeS are shown in Fig. 2. 

\subsection{Targeted Component}

Zeeman signatures are detected repeatedly in spectra of TC targets. The spectropolarimetric timeseries are
interpreted using several magnetic field modeling codes at our disposal. For those stars for which 
Stokes $V$ LSD profiles will be the primary model basis, modeling codes such as those of Donati et al. 
(2006) or Alecian et al. (2008b) will be employed.  For those stars for which the signal-to-noise ratio in 
individual spectral lines is sufficient to model the polarisation spectrum directly, we will employ the Invers10 Magnetic Doppler Imaging code to 
simultaneously model the magnetic field, surface abundance structures and pulsation velocity field 
(Piskunov \& Kochukhov 2002, Kochukhov et al. 2004). The resultant magnetic field models will be 
compared directly with the predictions of fossil and dynamo theory (e.g. Braithwaite 2006, 2007, Duez \& Mathis 2010, Arlt 2008).  

Diagnostics of the wind and magnetosphere (e.g. optical 
emission lines and their linear polarisation, UV line profiles, X-ray photometry and spectroscopy, radio 
flux variations, etc.) are being modeled using both the semi-analytic Rigidly-Rotating Magnetosphere 
approach, the Rigid-Field Hydrodynamics (Townsend et al. 2007) approach and full MHD simulations using the ZEUS 
code (e.g. Stone \& Norman 1992; ud Doula et al. 2008; Wade et al. in these proceedings; see also Fig. 4). 

\begin{figure}
\center
 \includegraphics[width=15cm]{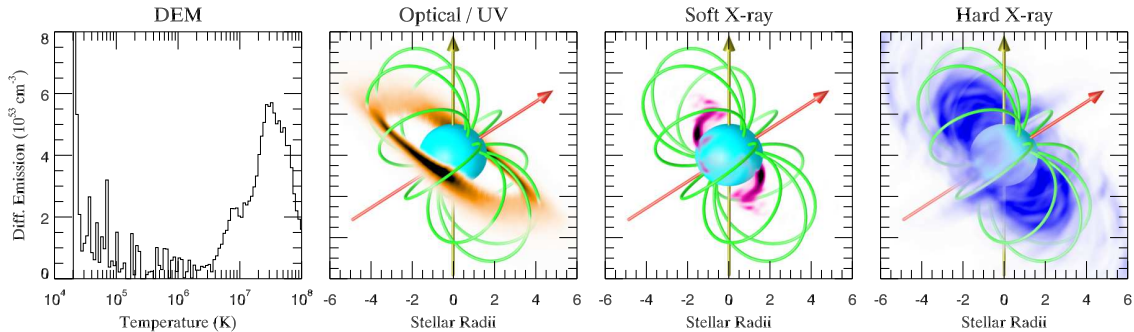}
  \caption{Example of the spectral and spatial emission properties of a rotating massive star magnetosphere modeled using Rigid Field Hydrodynamics (Townsend et al. 2007). The stellar rotation 
axis (vertical arrow) is oblique to the magnetic axis (inclined arrow), leading to complex plasma flow produced by radiative acceleration, Lorentz forces and centripetal acceleration. The consequent heated plasma distribution in the stellar magnetosphere (illustrated in colour/grey scale) shows broadband emission, and is highly structured both spatially and spectrally. Magnetically confined winds such as this are responsible for the X-ray emission and variability properties of some OB stars, and models such as this are being constructed for the MiMeS Targeted Component.
}\label{fig:wave}
\end{figure}

\section{Project status}

At the time of writing, MiMeS Large Programs have acquired over 1250 spectra of over 175 individual stars. About 45\% of the spectra correspond to the SC, while 65\% correspond to TC targets. Following their acquisition in Queued Service Observing mode at the CFHT, ESPaDOnS polarised spectra are immediately reduced by CFHT staff using the Upena pipeline feeding the Libre-Esprit reduction package and downloaded to the dedicated MiMeS Data Archive at the Canadian Astronomy Data Centre (CADC) in Victoria, Canada. Narval data are similarly reduced by TBL staff then downloaded to the MiMeS Collaboration's dedicated server at Observatoire de Paris, France. All reduced spectra are carefully normalized to the continuum using custom software tailored to hot stellar spectra. Each reduced spectrum is then subject to an immediate quick-look analysis to verify nominal resolving power, polarimetric performance and S/N. The quality of the $\sim 420$ SC spectra acquired to date is very high: the median peak SNR of the Stokes $I$ spectra is 1150 (per 1.8 km/s spectral pixel) and the median longitudinal field error bar ($1\sigma$) derived from the Stokes $V$ spectra is just 37 G. A typical SC spectrum is illustrated in Fig. 3.

Preliminary LSD profiles are extracted using our database of generic hot star line masks to perform an initial magnetic field diagnosis and further quality assurance. Ultimately, each spectrum will be processed by the MiMeS Massive Stars Pipeline (MSP; currently in production) to determine a variety of critical physical data for each observed target, in addition to the precision magnetic field diagnosis: effective temperature, surface gravity, mass, radius, age, variability characteristics, projected rotational velocity, radial velocity and binarity, and mass loss rate. These meta-data, in addition to the reduced high-quality spectra, will be uploaded for publication to the MiMeS Legacy Database\footnote{The MiMeS Project is undertaken within the context of the broader MagIcS (Magnetic Investigations of various Classes of Stars) collaboration, www.ast.obs-mip.fr/users/donati/magics).}.

A variety of MiMeS results are presented in these proceedings, and well as those of IAUS 272 (Active OB Stars). Here we provide a short description of a few notable results that have been obtained so far.

\section{Results}

\begin{figure}
\centering
\includegraphics[width=4in,angle=-90]{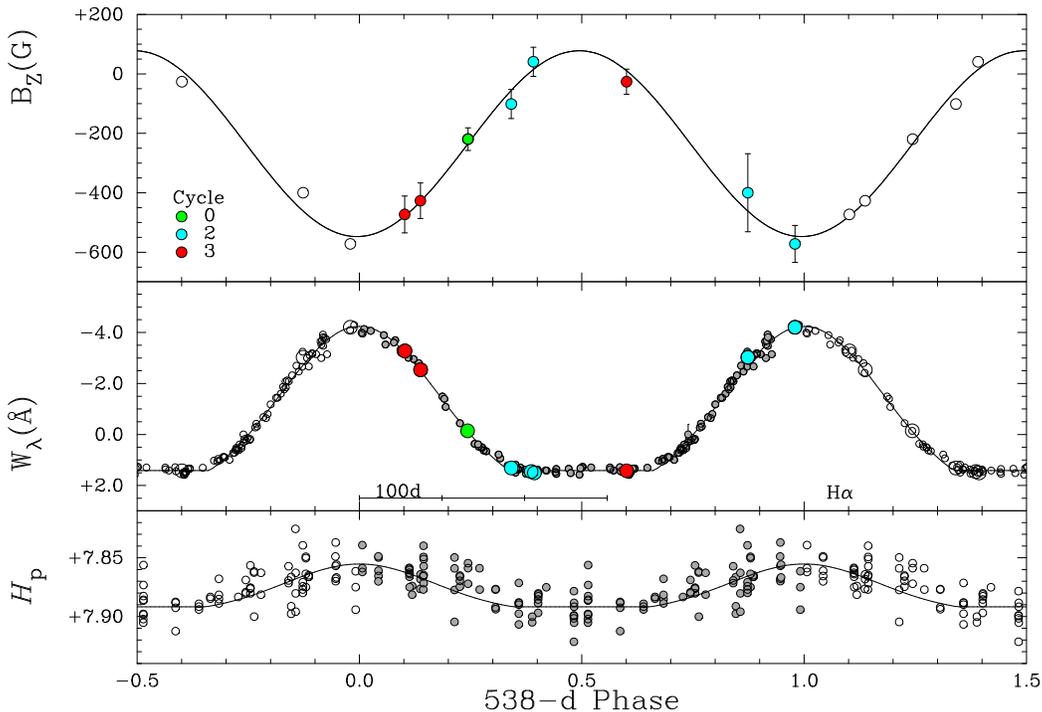}
\caption{Longitudinal field (top), H$\alpha$ EW (middle) and Hipparcos mag (bottom) of the Of?p star HD 191612, all phased according to the 537.6 d period. From Wade et al., in preparation. HD 191612 was the first Of?p star in which a magnetic field was detected (Donati et al. 2006). Subsequent MiMeS observations with ESPaDOnS (Wade et al., in prep) shown here confirm the existence of the field, and demonstrate the sinusoidal variability of the longitudinal field with the H$\alpha$ and photometric period of 537.6 d. The longitudinal field, H$\alpha$ and photometric extrema occur simultaneously implying a clear relationship between the magnetic field and the circumstellar envelope. }
\label{mag_geo}
\end{figure}

\subsection{Of?p stars}

The enigmatic Of?p stars are identified by a number of peculiar and outstanding observational properties. The classification was first introduced by Walborn (1972) according to the presence of C~{\sc iii} $\lambda 4650$ emission with a strength comparable to the neighbouring N~{\sc iii} lines. Well-studied Of?p stars are now known to exhibit recurrent, and apparently periodic, spectral variations (in Balmer, He~{\sc i}, C~{\sc iii} and Si~{\sc iii} lines) with periods ranging from days to decades, strong C~{\sc iii} $\lambda 4650$ in emission, narrow P Cygni or emission components in the Balmer lines and He~{\sc i} lines, and UV wind lines weaker than those of typical Of supergiants (see Naz\'e et al. 2010 and references therein). 

Only 5 Galactic Of?p stars are known (Walborn et al. 2010): HD 108, HD 148937, HD 191612, NGC 1624-2 and CPD$-28^{\rm o} 2561$. Three of these stars - HD 108, HD 148937 and HD 191612 - have been studied in detail. In recent years, HD 191612 was  carefully examined for the presence of magnetic fields (Donati et al. 2006), and was clearly detected. Recent observations, obtained chiefly within the context of the Magnetism in Massive Stars (MiMeS) Project (Martins et al. 2010; Wade et al., in prep) have furthermore detected magnetic fields in HD 108 and HD 148937, thereby confirming the view of Of?p stars as a class of slowly rotating, magnetic massive stars.

HD 191612 was the first Of?p star in which a magnetic field was detected (Donati et al. 2006). Subsequent MiMeS observations with ESPaDOnS (Wade et al., in prep) confirm the existence of the field, and demonstrate the sinusoidal variability of the longitudinal field with the H$\alpha$ and photometric period of 537.6 d. As shown in Fig. 5, the longitudinal field, H$\alpha$ and photometric extrema occur simultaneously when folded according to the 537.6 d period. This implies a clear relationship between the magnetic field and the circumstellar envelope. We interpret these observations in the context of the oblique rotator model, in which the stellar wind couples to the kilogauss dipolar magnetic field, generating a dense, structured magnetosphere, resulting in all observables varying according to the stellar rotation period.

\begin{figure}
\centering
\includegraphics[width=5in]{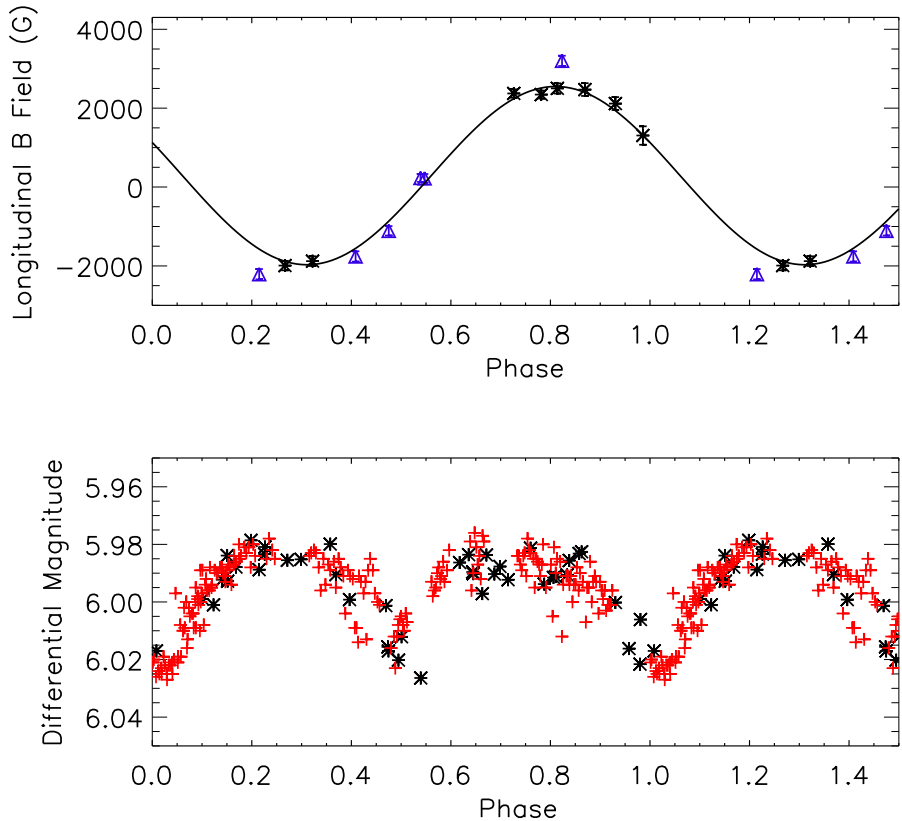}
\caption{Longitudinal field (top), and Hipparcos + SMARTS V magnitude photometry (bottom) of HR 7355 (Adapted from Oksala et al. 2010 and Rivinius et al. 2010).}
\label{mag_geo}
\end{figure}

\subsection{Rapidly rotating magnetic B-type stars}

Oksala et al. (2010) and Rivinius et al. (2010) simultaneously announced the detection of a strong longitudinal magnetic field in the rapidly-rotating He-strong main sequence B star HR 7355 (HD 182180). With a rotational period of $0.52144$ days, this star was identified as the most rapidly-rotating magnetic star known. It exhibits variable H$\alpha$ emission, and the longitudinal field, V magnitude and H$\alpha$ equivalent width all vary according to the derived period. 

Assuming a dipole magnetic geometry, the polar value of the magnetic field is 13-17 kG (Oksala et al. 2010). The approximate symmetry of the magnetic curve about $B_\ell=0$ implies that either the rotational axis inclination $i$ or the magnetic obliquity $\beta$ must be large (i.e. near 90$\degr$. This is consistent with the high $v\sin i$ ($300\pm 15$~km/s), which suggests that $\sin i$ must be rather close to unity. 

The photometric (brightness) light curve constructed from Hipparcos archival data and CTIO measurements shows two minima separated by 0.5 in rotational phase and occurring 0.25 cycles before/after the magnetic extrema. In combination with the H$\alpha$ emission, and in light of the behaviour of the prototypical magnetic He-strong star $\sigma$ Ori E, these properties suggest that HR 7355 is surrounded by a structured magnetosphere. The magnetospheric plasma is thought to have its origin in the stellar wind. Indeed, the rapid rotation of this star is expected to provide sufficient centrifugal support to wind plasma confined by the magnetic field that it is capable of forming a stable, long-lived and globally-structured magnetosphere. This is discussed in more detail by Wade et al. (these proceedings).

More recently, Grunhut et al. (2010) reported the detection of a strong magnetic field in the B2Ve star HR 5907 (HD 142184; see also Grunhut et al. in these proceedings). The spectrum of HR 5907 is remarkably similar to that of HR 7355, with very broad lines implying $v\sin i\sim 270$~km/s. Notwithstanding the somewhat lower $v\sin i$, the rotational period of this star inferred from Hipparcos photometry is $0.50831$~d - marginally shorter than HR 7355. Longitudinal field measurements obtained using He and metal lines show that only the negative magnetic hemisphere is presented to us, implying that the obliquity $\beta$ is rather small. HR 5907 also displays photometric variability and H$\alpha$ emission, again implying a structured magnetosphere.

Together, HR 7355, HR 5907 and $\sigma$~Ori E provide a fascinating sample for the testing of models of magnetospheric structure, such as the Rigidly Rotating Magnetosphere model developed by Townsend et al. (2005).

\subsection{Organisation and stability of fossil magnetic fields in stellar radiative zones}

Fossil fields are assumed to result from the trapping of the interstellar magnetic field during massive star formation or to be a remnant of a dynamo occurring during the pre-main sequence. If we assume that these initial fields are stochastic and turbulent, a significant challenge is to understand how these are converted into large-scale organized stable configurations as observed at the surface of MiMeS stars. This is closely related on the turbulent MHD relaxation mechanisms, in which an initial turbulent magnetic field is converted into a large-scale one - problems studied by the theoreticians of the MiMeS consortium. These investigators have studied the associated non-force free relaxed states of fields in radiative zones and they have presented the generalization of the known relation between magnetic energy and helicity in the stably stratified self-gravitating stellar case. Moreover, they have shown that the most probable state achieved after initial relaxation is a dipolar mixed configuration (with both a meridional and an azimuthal component) in the axisymmetric case that respects all known criteria to be stable found in previous numerical simulations (Duez \& Mathis 2010). 
	Earlier theoretical results on instability conditions of such fields have been verified using 3-D numerical simulations (Braithwaite 2006 \& 2007). It is therefore concluded that stable magnetic configurations should be of mixed type (with both poloidal and toroidal components) due to the stabilization of each component by the other (Braithwaite 2009). This is precisely the case of relaxed configurations obtained by numerical simulations and by semi-analytical methods. To study the stability of mixed configurations two methods can be used: the first semi-analytical one is based on the variational principle while the second one uses direct numerical simulations where the magnetic configuration is submitted to general perturbations. It is important to note that the second method is now intensively used because of the lack of general results from the analytical approach (which is very efficient to demonstrate instability, but not stability). This type of simulation has now been applied to relaxed configurations obtained by numerical simulations (Braithwaite 2009) and by semi-analytical methods (Duez, Braithwaite \& Mathis 2010) that have reached conclusions about the stability of those configurations, a major result for the magnetism of massive stars (and the resulting compact objects).
	MHD relaxation must now be studied in the general case, taking into account differential rotation and the obtained configurations have to be employed as initial conditions to study MHD transport processes in massive star interiors and their resulting evolution (Mathis \& Zahn 2005). Future work will involve the possibility of dynamo action in non-convective regions - a phenomenon to be studied in detail in the near future (Spruit 2002; Braithwaite 2006; Zahn, Brun \& Mathis 2007).

\section*{Acknowledgments}
The MiMeS CFHT Large Program (2008B-2012B) is supported by both Canadian and French Agencies, and was one of 4 such programs selected in early 2008 as a result of an extensive international expert peer review of many competing proposals. The MiMeS TBL Large Program (2010B-2012B) was allocated in the context of the competitive French Time Allocation process.

Based on observations obtained at the Canada-France-Hawaii Telescope (CFHT) which is operated by the National Research Council of Canada, the Institut National des Sciences de l'Univers of the Centre National de la Recherche Scientifique of France, and the University of Hawaii. Also based on observations obtained at the Bernard Lyot Telescope (TBL, Pic du Midi, France) of the Midi-Pyr\'en\'ees Observatory, which is operated by the Institut National des Sciences de l'Univers of the Centre National de la Recherche Scientifique of France.

The MiMeS Data Access Pages are powered by software developed by the CADC, and contains data and meta-data provided by the CFH Telescope.

\end{document}